\documentclass[sigconf]{acmart}

\setcopyright{none}  
\renewcommand\footnotetextcopyrightpermission[1]{}  

\acmConference[Augmented Educators and AI(CHI 2025 Workshop)]{Augmented Educators and AI: Shaping the Future of Human-AI Collaboration in Learning on CHI 2025 Workshop}{April 26,2025}{Yokohama, JAPAN}

\begin{document}

\title[LLM-Based Teachable Agent for Learning Gains and Cognitive Load in Music Education]{Exploring the Impact of an LLM-Powered  Teachable Agent on Learning Gains and Cognitive Load in Music Education}

\author{Lingxi Jin}
\email{jinlingxi@ewha.ac.kr}
\affiliation{%
  \institution{Ewha Womans University}
  \country{South Korea}
}

\author{Baicheng Lin}
\email{linbaicheng@sju.ac.kr}
\affiliation{%
  \institution{Sejong University}
  \country{South Korea}
}

\author{Mengze Hong}
\email{mengze.hong@connect.polyu.hk}
\affiliation{%
  \institution{Hong Kong Polytechnic University}
  \country{Hong Kong, China}
}

\author{Kun Zhang}
\email{kun2.zhang@live.uwe.ac.uk}
\affiliation{%
  \institution{University of the West of England}
  \country{United Kingdom}
}

\author{Hyo-Jeong So}
\email{hyojeongso@ewha.ac.kr}
\affiliation{%
  \institution{Ewha Womans University}
  \country{South Korea}
}
\authornote{Corresponding to: hyojeongso@ewha.ac.kr}

\renewcommand{\shortauthors}{Jin et al.}

\begin{abstract}
This study examines the impact of an LLM-powered teachable agent, grounded in the Learning by Teaching (LBT) pedagogy, on students’ music theory learning and cognitive load. The participants were 28 Chinese university students with prior music instrumental experiences. In an online experiment, they were assigned to either an experimental group, which engaged in music analysis with the teachable agent, or a control group, which conducted self-directed analysis using instructional materials. Findings indicate that students in the experimental group achieved significantly higher post-test scores than those in the control group. Additionally, they reported lower cognitive load, suggesting that the teachable agent effectively reduced the cognitive demands of music analysis tasks. These results highlight the potential of AI-driven scaffolding based on LBT principles to enhance music theory education, supporting teachers in delivering theory-oriented instruction while fostering students’ self-directed learning skills.
\end{abstract}

\keywords{LLM-based teachable agent, music education, learning by teaching, cognitive load, learning gains, AI-assisted learning}

\maketitle

{\small
\noindent ©  This paper was adapted for the \textit{CHI 2025 Workshop on Augmented Educators and AI: Shaping the Future of Human and AI Cooperation in Learning},
held in Yokohama, Japan on April 26, 2025. This work is licensed under the Creative Commons Attribution 4.0 International License (CC BY 4.0).
}

\section{INTRODUCTION}
Learning by Teaching (LBT) is a pedagogical approach that strengthens knowledge acquisition by requiring students to explain concepts \cite{biswas2005learning}. While peer teaching has been widely implemented in traditional classrooms, logistical constraints often limit its scalability. Teachable Agents (TAs), AI-driven virtual learners, help mitigate this challenge by prompting students to articulate, refine, and validate their understanding rather than passively receiving information. Previous studies have shown that students interacting with LLM-powered TAs demonstrate greater learning gains and improved self-regulation skills \cite{chen2024learning} compared to those receiving traditional instruction. Unlike rule-based TAs \cite{jin2024teach, matsuda2011learning}, LLM-powered TAs facilitate dynamic, iterative dialogue, guiding students through structured reasoning \cite{jin2024teach}. While extensively studied in STEM education \cite{chen2024learning} \cite{jin2024teach} \cite{song2024students}, the application of LLM-powered TAs in music education remains largely unexplored. Given the complex cognitive demands of music theory learning, this study introduces Chat Melody, an LLM-based TA designed to complement rather than replace traditional instructor-centered pedagogy. Chat Melody serves as an interactive learning companion, encouraging students to explain, justify, and refine their understanding through structured dialogue. This study aims to address the following research questions:

\begin{itemize}
    \item RQ1: How does the teachable agent Chat Melody impact students’ knowledge gains in music theory?
    \item RQ2: How does the teachable agent Chat Melody influence students’ cognitive load during music analysis?
\end{itemize}

\section{BACKGROUND}
\subsection{Learning by Teaching and Teachable Agents }

Learning by Teaching (LBT) is an instructional strategy that fosters knowledge acquisition and metacognitive awareness by requiring students to articulate, explain, and refine their understanding through the process of teaching others \cite{biswas2005learning}. Rooted in social constructivist theory, LBT emphasizes active engagement in structured dialogue and problem-solving, allowing learners to develop deeper conceptual understanding through knowledge co-construction. The process of exploring, explaining, and reflecting ideas enhances conceptual understanding \cite{davis2006characterizing}. However, without structured support, learners often engage in knowledge-telling \cite{roscoe2007understanding}, simply recalling and transmitting information without critically reflecting on or reorganizing their understanding \cite{pareto2011teachable}. 

To address this limitation, Teachable Agents (TAs) as AI-driven virtual tutees have been introduced to support LBT environments. 

Unlike traditional intelligent tutoring systems that function as directive knowledge providers, TAs act as cognitive scaffolds, prompting learners to actively explain concepts, correct misconceptions, and refine their reasoning \cite{biswas2005learning}. Empirical studies indicate that TAs can enhance students’ conceptual retention \cite{zhao2012learning}, learning engagement, and self-regulation by simulating interactive peer-teaching experiences \cite{chin2010preparing} \cite{pareto2011teachable}. However, early rule-based TAs were constrained by predefined and rigid interaction structures, limiting their ability to support flexible and adaptive learning experiences \cite{johnson2018pedagogical, matsuda2011learning}. These systems primarily relied on non-interactive knowledge representations such as concept maps where students directly edited the agent’s knowledge state instead of engaging in open-ended discussions \cite{chase2009teachable}. Recent advancements in LLM-powered TAs have addressed this limitation by enabling context-sensitive, free-form dialogue, allowing for more human-like and adaptive learning interactions \cite{jin2024teach}.

\subsection{LLM-Powered Teachable Agents to Augment Human Teachers}

Studies on LLM-powered teachable agents have demonstrated their effectiveness in enhancing learning gains across multiple domains. In programming education \cite{chen2024learning, jin2024teach}, students who engaged in teaching an AI-driven agent exhibited higher test scores and problem-solving proficiency compared to those following traditional instructional methods. By applying structured interaction models, such as Gall’s help-seeking framework, these AI-driven systems encourage learners to internalize concepts and progress through deliberate cognitive stages rather than merely absorbing AI-generated responses. One key advantage of LLM-powered TAs is their ability to provide real-time, personalized feedback, ensuring that students receive adaptive scaffolding suited to their individual learning needs. By allowing learners to actively teach AI-driven agents, these systems enhance long-term knowledge retention and metacognitive self-regulation, reinforcing deep learning rather than rote memorization.

While prior studies underscore the learning benefits of LLM-based TAs in STEM fields, their application in music education remains underexplored. Music theory serves as the foundation of music education, providing learners with the essential framework for analyzing, interpreting, and composing music \cite{fong2010music, ashford1966use} \cite{chong2019teaching}. Traditional teacher-centered instruction remains dominant, with students primarily engaging in lectures, guided exercises, and score analysis \cite{ashford1966use}. However, this approach presents significant challenges, particularly in large-class settings, where instructors struggle to provide timely, individualized feedback \cite{wible2001web}. As a result, many students rely on rote memorization rather than engaging in deep conceptual reasoning, limiting their ability to apply theoretical knowledge flexibly. The lack of immediate, adaptive feedback in traditional instruction creates barriers to effective learning, particularly in complex problem-solving tasks where students must synthesize harmonic structures, rhythmic patterns, and compositional techniques \cite{ashford1966use}. Integrating LLM-powered TAs into music education offers a potential solution to this issue. These agents provide real-time, interactive feedback, encouraging students to engage in iterative self-explanation, reasoning, and correction of misconceptions rather than passively absorbing information.

\section{THE DESIGN OF AN LLM-BASED TEACHABLE AGENTS}
Our research team initially developed this LLM-powered multimodal tutoring system to assist music learners in solving music analysis tasks \cite{jinexploring}. The system incorporates three core interactive components (see Figure 1): 1) problem statement area presenting structured music analysis tasks, prompting learners to evaluate harmonic progressions, melodic structures, and expressive elements; 2) interactive music sheet display, allowing learners to click on specific notes, triggering dynamic highlighting, zooming, and MIDI playback to reinforce connections between notation and auditory perception; and 3) a teachable agent powered by OpenAI GPT-4, designed to simulate a novice learner, ensuring structured dialogues through instruction tuning. The dialogue example originally in Chinese was translated into English by the research team.

To guide students through structured help-seeking interactions, the system follows the five-stage help-seeking process \cite{nelson1981help}: (1) recognizing the need for help, (2) deciding to seek assistance, (3) identifying a source, (4) employing strategies, and (5) responding to guidance. For example, when analyzing a harmonic structure in Mozart’s Piano Sonata No. 5 in G major, K. 283 (Movement 2), the teachable agent may prompt reflection rather than supplying immediate answers. Instead of directly identifying a chord, it asks, “I see E, G, C, and G in the right hand—what does this chord represent?” This encourages students to analyze the given notes, leading them to recognize a C major chord (IV in G major). The agent then guides further reasoning by questioning the influence of the bass note, prompting learners to identify the cadential six-four (IV6/4 → V → I) harmonic resolution. 

\begin{figure*}
    \centering
    \includegraphics[width=0.63\linewidth]{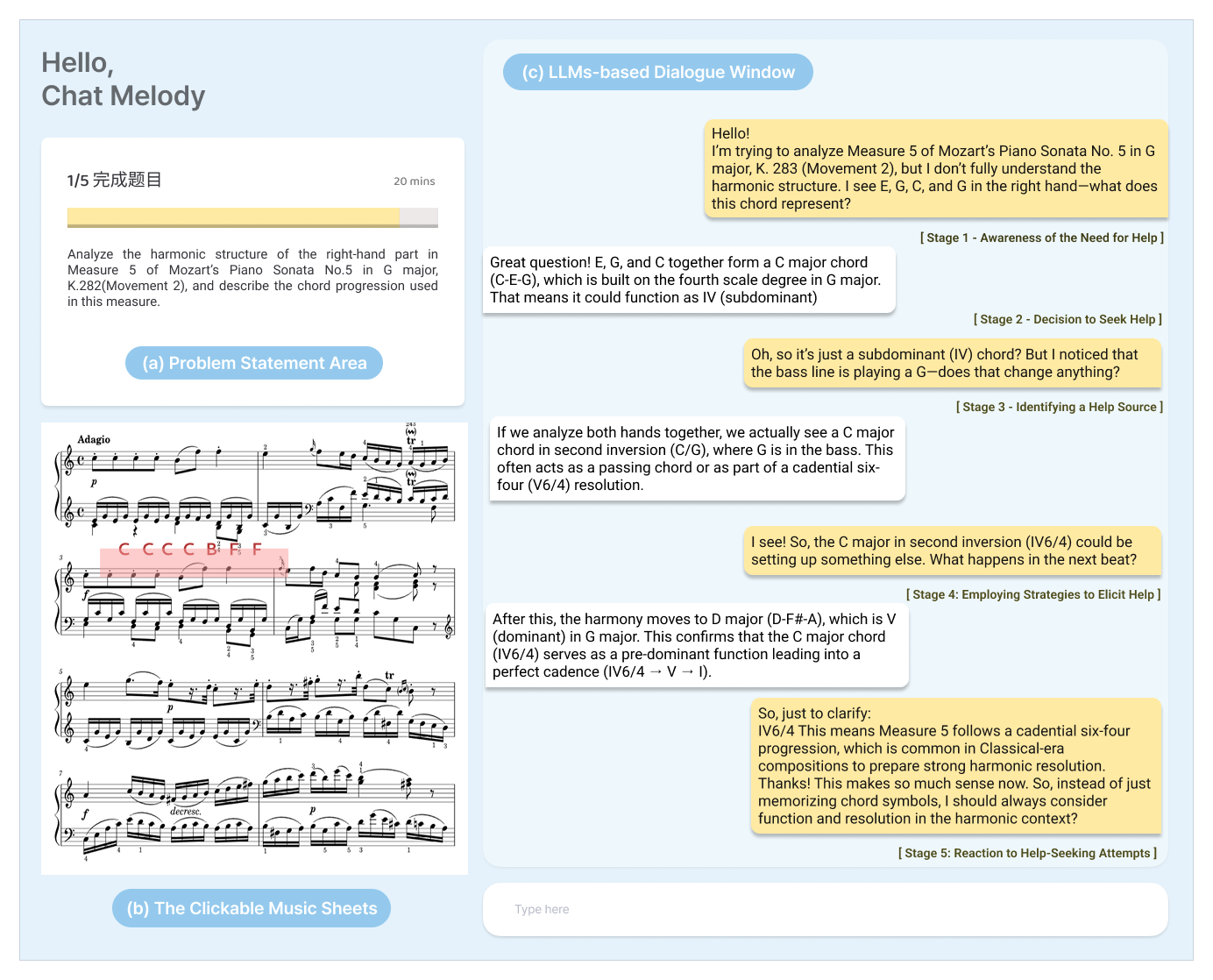}
    \caption{ Interface design of the LLM-powered Teachable Agent \textit{Chat Melody}}
    \label{fig:enter-label}
\end{figure*}

\section{METHODS}

\subsection{Participants \& Experiment Design}
We conducted an experiment to evaluate the impact of an LLM-powered TA on music theory learning and cognitive load. The study was conducted entirely online and included 28 Chinese university students (18 females, 10 males) aged 19 to 29 years ($M = 24.29$, $SD = 2.62$), all of whom had prior instrumental learning experience and foundational knowledge in music theory. Participants were assigned to either the experimental group ($n = 16$), which engaged with the LLM-based TA, or the control group ($n = 16$), which received teacher-guided instruction. However, four participants from the control group withdrew before completing the study, leaving a final sample of 12 students in the control group. 

The study followed a three-phase experimental procedure: pre-test, instructional phase, and post-test. Before the instructional phase, all participants completed a pre-test assessing their knowledge of music notation, scales, and harmonic structures to establish baseline knowledge levels. The instructional phase spanned two days and was delivered via Zoom, with all participants attending the same sessions, ensuring that both groups received the same foundational instruction. On the first day, participants learned about music notation and interpretation, covering topics such as note values, rhythm, articulation, scales, and key signatures, reinforced through guided exercises and listening activities. On the second day, participants explored musical structures, with a focus on sonata form, a key component in analyzing Mozart’s Piano Sonata No. 5 in G Major, K. 283, Movement 2. They examined thematic development, modulation, and harmonic progressions within the exposition, development, and recapitulation sections of the piece. 

Following the instructional phase, participants completed a 60-minute music analysis task for the post-test, applying their theoretical knowledge to Mozart’s Piano Sonata No. 5 in G Major, K. 283, Movement 2. While all participants had received the same instruction, their approach to the analysis task differed depending on their assigned condition. The experimental group engaged in structured dialogues with the TA, using interactive reasoning to articulate concepts, justify interpretations, and refine analyses. The TA did not provide direct answers but instead prompted learners to identify harmonic structures, justify their reasoning, and refine their interpretations through iterative feedback. In contrast, the control group performed the same analysis independently, relying solely on instructional materials and recorded lecture videos. To ensure consistency and prevent external resource use, participants shared their screens via Zoom, and a researcher monitored their progress. After the task, all participants completed a cognitive load assessment \cite{paas2016cognitive}, measuring mental load (content complexity) and mental effort (cognitive resources required).

\subsection{Measurement}

\subsubsection{Knowledge set}

To assess learning gains in music theory, we administered pre-test and post-test assessments. The test was designed by an experienced music theory lecturer, ensuring content validity and alignment with the learning objectives. The knowledge test consisted of 15 multiple-choice questions, covering core topics such as notation, rhythm, harmonic structures, and form analysis. Each question was carefully designed to assess different levels of cognitive processing, ranging from basic recall to analytical reasoning. The test included 5 easy, 5 medium, and 5 difficult questions, with each question worth 1 point, resulting in a maximum score of 15 points. The same test format was applied to both the pre-test and post-test, with variations in the question content to prevent memorization effects. Scores were standardized to enable a comparative analysis of knowledge improvement. 

\subsubsection{Cognitive load}
The cognitive load questionnaire used in this study was adapted from the established measures \cite{paas2016cognitive} and included eight items on a six-point Likert scale. The questionnaire assessed two key dimensions: mental load (five items) and mental effort (three items), with Cronbach's alpha values of .88 and .77, respectively, indicating high internal reliability. For example, mental load items asked participants to rate the difficulty of the learning content and the effort required to complete tasks, such as ``I had to put a lot of effort into answering the questions.'' Similarly, mental effort items evaluated the cognitive demands of the instructional approach, such as ``I had to put in significant effort to complete the learning tasks.''

\begin{table*}[!t]
\centering
\small
\caption{t-Test result of the pre-test scores}
\begin{tabular}{llccccc}
\hline
        & \textbf{Group}           & \textbf{N} & \textbf{Mean} & \textbf{S.D.} & \textbf{\textit{t}} & \textbf{\textit{p}} \\ \hline
\textbf{Pre-test} & Experimental group & 16         & 48.58        & 11.14        & .86        & .369       \\  
        & Control group      & 12         & 51.89        & 8.33         &             &            \\ \hline
\end{tabular}
\end{table*}

\begin{table*}[!t]
\centering
\caption{Descriptive data and ANCOVA of the post-test results}
\begin{tabular}{lcccccccc}
\hline
\textbf{Group}           & \textbf{N} & \textbf{Mean} & \textbf{S.D.} & \textbf{Adjusted Mean} & \textbf{S.E.} & \textbf{F} & \textbf{P} & $\mathbf{\eta^2}$ \\ \hline
Experimental group & 16         & 57.30        & 6.72         & 59.65            & 1.69        & 19.98      & \textless.001 & 0.35      \\ 
Control group      & 12         & 50.07        & 8.30         & 46.94            & 2.00        &            &              &           \\ \hline
\end{tabular}
\end{table*}

\section{RESULTS}

\subsection{How does the teachable agent impact students' knowledge learning gains in music theory?}

Before the instructional phase, an independent t-test was performed on pre-test scores to ensure no significant differences in prior knowledge between the experimental group and the control group. The results confirmed comparable baseline knowledge between the two groups ($t = .86$, $p > .05$). To assess the effectiveness of the TA, an ANCOVA was conducted using post-test scores as the dependent variable and pre-test scores as the covariate. The adjusted mean post-test scores were 59.65 ($SE = 1.69$) for the experimental group and 46.94 ($SE = 2.00$) for the control group. The ANCOVA results indicated a statistically significant difference between the groups ($F = 19.98$, $p < .001$). Levene’s test confirmed that the assumption of equal variances was met ($F = .59$, $p = .450$), validating the ANCOVA analysis.

\subsection{How does the teachable agent influence students’ cognitive load during music analysis?}

Since learning with an LLM-based TA was a new experience for the participants, it was essential to investigate their cognitive load during the learning process. As shown in Table \ref{tab: table3}, students in the experimental group who used Chat Melody reported a significantly lower overall cognitive load ($M = 2.82$, $SD = .60$) compared to those in the control group ($M = 3.43$, $SD = .44$), with a t-test confirming a significant difference ($t = 2.97$, $p <.05$).

\begin{table}[ht]
\centering
\caption{\textit{t}-Test result of the cognitive load on the post-questionnaire scores of the two groups}
\begin{tabular}{lccccccc}
\hline
\textbf{Group}           & \textbf{N} & \textbf{Mean} & \textbf{S.D.} & \textbf{\textit{t}} & \textbf{\textit{p}} \\ \hline
Experimental group & 16         & 2.82        & .60         & 2.97* & .006      \\ 
Control group      & 12         & 3.43        & .44         &             &           \\ \hline
\end{tabular}
\begin{flushleft}
*$p < .05$
\end{flushleft}
\label{tab: table3}
\end{table}

To further analyze cognitive load, we examined its two dimensions: mental load and mental effort (see Table \ref{tab: table4}). While the experimental group experienced slightly lower mental load ($M = 2.81$, $SD = .84$) than the control group ($M = 3.32$, $SD = .44$), the difference was not statistically significant ($t = 1.89$, $p >.05$), indicating that both groups perceived the task complexity similarly. However, a significant difference was observed in mental effort, with the experimental group ($M = 2.83$, $SD = .67$) reporting significantly lower effort compared to the control group ($M = 3.61$, $SD = 1.00$; $t = 2.47$, $p < .05$). This suggests that Chat Melody effectively reduced the cognitive demands placed on learners by structuring their reasoning process, making it easier for them to process and apply theoretical concepts. 

\begin{table}[ht]
\centering
\small
\caption{\textit{t}-Test result of the cognitive load dimensions on the post-questionnaire scores of the two groups}
\resizebox{\columnwidth}{!}{%
\begin{tabular}{l l c c c c c}
\hline
               & \textbf{Group}           & \textbf{N} & \textbf{Mean} & \textbf{S.D.} & \textbf{\textit{t}} & \textbf{\textit{p}} \\ \hline
\textbf{Mental Load} & Experimental group & 16         & 2.81        & .84         & 1.89 & .070      \\ 
               & Control group      & 12         & 3.32        & .44         &      &        \\ \hline
\textbf{Mental Effort} & Experimental group & 16         & 2.83        & .67         & 2.47 & .021      \\ 
               & Control group      & 12         & 3.61        & 1.00        &      &        \\ \hline
\end{tabular}}
\begin{flushleft}
*$p < .05$
\end{flushleft}
\label{tab: table4}
\end{table}

\section{DISCUSSION AND CONCLUSIONS}

This study examined the impact of an LLM-powered TA Chat Melody on students’ music theory learning and cognitive load, by integrating the Learning by Teaching (LBT) approach to promote structured reasoning and self-explanation. Overall, the findings indicate that students who engaged with Chat Melody demonstrated significantly higher learning gains compared to those relying solely on self-directed problem solving. This finding supports previous research on the effectiveness of interactive scaffolding in facilitating deeper conceptual understanding \cite{matsuda2011learning}. The teachable agent encouraged students to articulate their reasoning, justify theoretical interpretations, and refine their analyses through iterative dialogue, reinforcing knowledge retention and enhancing their ability to apply theoretical concepts. 

In addition to learning gains, this study explored cognitive load in AI-assisted music learning. The results showed that students using Chat Melody experienced significantly lower cognitive load, particularly in mental effort. While both groups faced similar levels of mental load due to the inherent complexity of music theory learning, the experimental group reported less effort to process and apply their knowledge. This suggests that the structured guidance provided by the teachable agent effectively reduced cognitive demands, aligning with cognitive load theory, which emphasizes the importance of instructional designs that optimize cognitive resources for meaningful learning.

This study has some limitations to be considered in future research. The small sample size may limit the generalizability of the findings, indicating the need for studies involving larger and more diverse participant groups. Additionally, this study did not measure long-term retention, suggesting a need for delayed post-tests in future research. Despite these limitations, this study provides valuable insights into the role of AI-powered TAs in music education. In the absence of immediate teacher feedback, these agents can serve as cognitive scaffolds, guiding learners through structured reasoning and fostering deeper understanding. This suggests that interactive TAs not only support self-directed learning but also offer a meaningful complement to traditional music theory instruction, empowering human teachers to provided more personalized and interactive experiences to learners.

\bibliographystyle{ACM-Reference-Format}
\bibliography{sample-base}

\end{document}